# Impact of inclusive electron ion collider data on collinear parton distributions

Néstor Armesto,[1] Thomas Cridge[2,*], Francesco Giuli,[3] Lucian Harland-Lang,[4] Paul Newman,[5] Barak Schmookler,[6] Robert Thorne,[4] and Katarzyna Wichmann[2]

[1]*Departamento de Física de Partículas and IGFAE, Universidade de Santiago de Compostela, 15782 Santiago de Compostela, Galicia, Spain*
[2]*Deutsches Elektronen-Synchrotron DESY, Germany*
[3]*CERN, CH-1211 Geneva, Switzerland*
[4]*Department of Physics and Astronomy, University College, London, WC1E 6BT, United Kingdom*
[5]*School of Physics and Astronomy, University of Birmingham, B15 2TT, United Kingdom*
[6]*Department of Physics and Astronomy, University of California, Riverside, California 92521, USA*

A study is presented of the impact of updated simulated inclusive electron ion collider deep inelastic scattering data on the determination of the proton and nuclear parton distribution functions (PDFs) at next-to-next-to-leading and next-to-leading order in QCD, respectively. The influence on the proton PDFs is evaluated relative to the HERAPDF2.0 set, which uses inclusive HERA data only, and also relative to the global fitting approach of the MSHT20 PDFs. The impact on nuclear PDFs is assessed relative to the EPPS21 global fit and is presented in terms of nuclear modification ratios. For all cases studied, significant improvements in the PDF uncertainties are observed for several parton species. The most striking impact occurs for the nuclear PDFs in general and for the region of high Bjorken $x$ in the proton PDFs, particularly for the valence quark distributions.

## I. INTRODUCTION

Parton distribution functions (PDFs) are essential tools for understanding experimental data from collider experiments, and a necessary part of any theory prediction for hard hadronic interactions. Since the PDF dependence on Bjorken $x$ cannot be calculated perturbatively, it has to be extracted from experimental measurements. deep inelastic scattering (DIS) data from the HERA $ep$ collider are at the core of current proton PDF extractions. In different approaches, they are used either as the sole ingredients [1] or else in global fits [2–7] together with fixed target DIS data and measurements of PDF-sensitive processes at the Tevatron and LHC hadron colliders. Nuclear parton distribution functions (nPDFs) describe the behavior of bound partons in the nucleus. Since there has yet to be an electron-nucleus collider, the standard approach is to combine fixed target DIS and Drell-Yan data with hard process measurements ($\pi^0$, $D$, electro-weak boson and jet production) from proton-nucleus collisions at RHIC and the LHC, in global analyses [8–12], see [13,14] for recent reviews.

The Electron Ion Collider (EIC) is expected to begin operations around 2030 at Brookhaven National Laboratory in the USA, with a broad scientific program of enhancing our knowledge of nucleon and nuclear structure and dynamics. It will provide very precise $ep$ DIS data in a complementary kinematic region to that of HERA. The prospect of EIC data at large $x$ and intermediate $Q^2$ is of particular interest, since existing large $x$ DIS data from HERA are limited in their experimental precision, while fixed target data are subject to nuclear corrections and other theoretical complications. Many of these fixed target data are at low $W^2$ where higher twist corrections are known to be enhanced and where target-mass corrections can also be necessary. The impact of the EIC on proton PDFs has been studied previously in [15–19]. In the present work we analyse the impact of EIC pseudodata simulated with updated energy setups, luminosities and more realistic detector simulation [20] relative to previous studies. We consider both DIS-only and global fits, analyzing the impact of this data in the context of the HERAPDF2.0 and MSHT PDF frameworks for the first time, and study the impact on parton-parton luminosities and selected observables in hadronic collisions. In addition, in this study we also examine for the first time the sensitivity

---

*thomas.cridge@desy.de

of inclusive measurements at the EIC on resummation of small-$x$ logarithms.

As the world's first $eA$ collider, the EIC will explore nuclear structure to an unprecedented level of detail up to the heaviest nuclei, in particular providing the first sensitivity from a DIS experiment in the low $x$ kinematic regime. The impact of the EIC on nPDFs has previously been considered in [15,16,19,21,22]. Compared to those studies that include EIC pseudodata in global fits containing data from different collision systems and nuclei, in the present work we perform fits to pseudodata for a single nucleus (Au) with the same techniques used for the proton. We show that the extraction of nPDFs for a single nucleus is possible using only EIC pseudodata and, therefore, with reduced uncertainties and parametrization bias compared to global fits. As for proton data, the energy setups, integrated luminosities and detector simulation [20] have been updated[1] and can be considered more realistic.

In this work we study the impact of the EIC on proton and nuclear PDFs using simulated EIC inclusive cross sections for neutral current (NC) and charged current (CC) DIS, with kinematic range, resolutions and uncertainties derived from a realistic detector design [20].[2] The uncertainties obtained in fits with parton evolution governed by the DGLAP equations [30–34] at next-to-next-to-leading order (NNLO) [35–37] are compared between baselines given by existing PDF sets and projections with the additional inclusion of simulated EIC data. A closely related study that assesses the EIC impact on the strong coupling $\alpha_S$ can be found in [38].

## II. SIMULATED EIC DATA

The detailed experimental apparatus to be deployed at the EIC is currently under intense development. However, the broad requirements to realize the science programme are well established [15], including the kinematic range and level of performance expected for inclusive DIS. For the purpose of this work, the simulated EIC data are taken from the studies performed in the framework of the ATHENA detector proposal [20]. Since the proposal publication, ATHENA concepts have been combined with those from the ECCE proposal [39] in the framework of a new ePIC collaboration, while discussions about a second complementary detector are also ongoing. The results here are not expected to differ significantly for any general purpose EIC detector and are more up-to-date than previous studies.

---

[1]Some preliminary partial results of our study were shown in [20]. They have been largely extended and updated here.
[2]While we restrict our study to inclusive observables, we note that the impact of less inclusive ones on (n)PDF determination has been studied elsewhere, e.g., charm production [21,23], jets in DIS [24,25] and photoproduction [26–28] or charm-tagged jets [29].

TABLE I. Beam energies, center-of-mass energies and annual integrated luminosities of the different configurations considered for the EIC.

| $e$-beam energy (GeV) | $p$-beam energy (GeV) | $\sqrt{s}$ (GeV) | Integrated lumi (fb$^{-1}$) |
|---|---|---|---|
| 18 | 275 | 141 | 15.4 |
| 10 | 275 | 105 | 100 |
| 10 | 100 | 63 | 79.0 |
| 5 | 100 | 45 | 61.0 |
| 5 | 41 | 29 | 4.4 |

Simulated neutral current (NC) EIC $ep$ measurements ("pseudodata"), are produced with integrated luminosities corresponding to expectations for one year of data taking with each of the five different currently planned beam configurations. The details are outlined in Tab. I. Charged current (CC) pseudodata are also available though only at the highest $\sqrt{s}$. For the NC pseudodata, a grid is produced with five logarithmically spaced $x$ and $Q^2$ values per decade over the range[3] $0.001 < y < 0.95$, which is well-justified by the expected resolutions.

The central values of the pseudodata are taken from predictions using the HERAPDF2.0NNLO [1] and MSHT20 NLO [3] PDF sets (for Secs. III A and III B, respectively). In the latter case, the NLO set (and matching NLO theory) is taken in order to inject some pseudodata/theory disagreement, as discussed further in Sec. III B. The pseudodata points are randomly smeared based on Gaussian distributions with standard deviations given by the projected uncertainties as estimated by the ATHENA collaboration [20,40]. The simulated systematic errors are based on experience from HERA and further considerations in [15] and are rather conservative in the context of the modern detector technologies and large datasets expected at the EIC. Most data points have a point-to-point uncorrelated systematic uncertainty of 1.9%, extending to 2.75% for $y < 0.01$. A normalization uncertainty of 3.4% is also ascribed; it is taken as fully correlated between data at each $\sqrt{s}$, and fully uncorrelated between datasets with different $\sqrt{s}$. For the purposes of the QCD fits, the point-to-point systematic uncertainties are added in quadrature with the statistical uncertainties and the normalization uncertainties are treated as nuisance parameters, as in [1].

The locations in the $(x, Q^2)$ kinematic plane of the HERA data and EIC pseudodata used in this analysis are shown in Fig. 1. The EIC pseudodata overlap in their coverage with the HERA data and extend the kinematic reach to the highest $x$ values in the intermediate $Q^2$ region.

The large $x$ HERA data are relatively imprecise due to the kinematic correlation with large $Q^2$, the $1/Q^4$ photon propagator term in the cross section and the modest integrated luminosity ($\sim 0.5$ fb$^{-1}$ per experiment). The large $x$ region in global fits is therefore constrained to a large extent by measurements from fixed target experiments, e.g., BCDMS and NMC [41,42]. However, there are uncertainties in the theoretical description of the fixed target data due to their low hadronic final state invariant masses,[4] values where it becomes difficult to disentangle perturbative corrections from powerlike effects. The EIC is thus particularly promising in the high $x$ region, where it is expected to provide data that are both high precision and theoretically clean.

---

[3]Here, $y$ is the usual inelasticity variable, $y = Q^2/(sx)$.

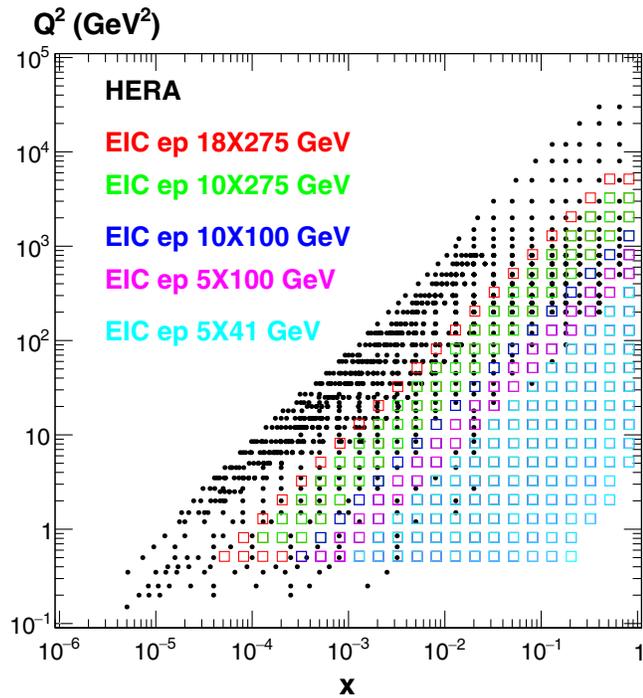

FIG. 1. The locations in the $(x, Q^2)$ plane of the HERA (black solid points) and EIC (open symbols) $ep$ neutral current inclusive DIS data points included in the analysis.

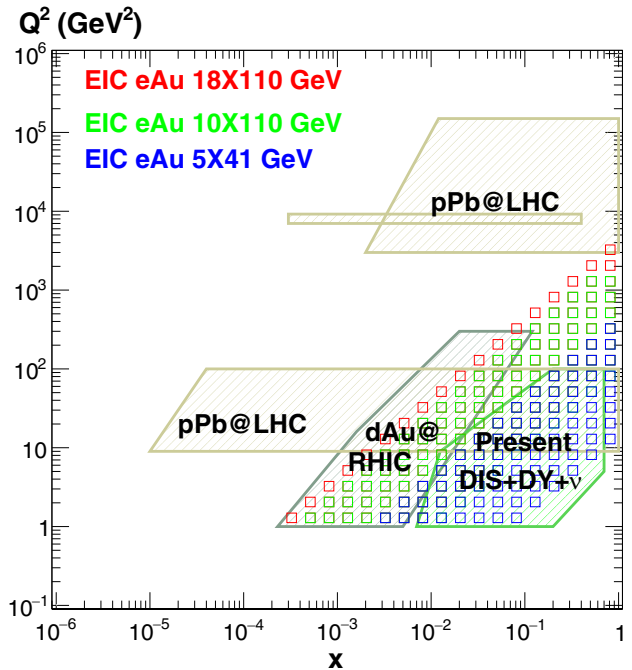

FIG. 2. The locations in the $(x, Q^2)$ plane of the $eAu$ EIC neutral current inclusive DIS data points included in the analysis (open symbols), compared to the region (hatched areas) covered at present by DIS and Drell-Yan fixed target experiments on nuclear targets, and by dijet, electro-weak boson and D-meson production in $pPb$ collisions at the LHC.

$eA$ pseudodata were produced analogously, considering the nucleus to be Au, and per-nucleon integrated luminosities of 4.4 fb$^{-1}$, 79 fb$^{-1}$ and 79 fb$^{-1}$ for $5 \times 41$ GeV, $10 \times 110$ GeV and $18 \times 110$ GeV, respectively. The locations in the $(x, Q^2)$ kinematic plane of the EIC pseudodata used in this analysis are shown in Fig. 2, together with shaded areas representing the regions presently covered and considered in existing global nPDF fits [13,14]. Note that we are interested in the uncertainties while the central values are irrelevant for this study. Therefore, the same PDF set HERAPDF2.0NNLO [1] used for the proton is employed for $eA$, corresponding to a central value of the nuclear modification factor (defined as the ratio of each parton density in a proton bound inside a nucleus to that in a free proton) equal to 1.

### III. EIC IMPACT ON PROTON PDFs

#### A. Comparison with HERA-only PDFs

The results presented in this section are obtained from global QCD fits at NNLO, performed in the HERAPDF2.0 framework [1] using xFitter, an open source QCD fit platform [43]. Fits with identical configurations are performed to HERA data only, corresponding to HERAPDF2.0NNLO in [1], and also with the additional inclusion of the simulated EIC pseudodata described in Sec. II. To avoid regions that may be strongly affected by higher twist or resummation effects, a cut on the squared hadronic final state invariant mass, $W^2 = Q^2(1-x)/x > 10$ GeV$^2$ is included for the EIC data. No such cut was required in the HERAPDF2.0NNLO fit as the kinematic range of the data included there is such that $W^2 \gtrsim 270$ GeV$^2$. The central values of the PDFs with and without the EIC pseudodata coincide by construction, so the uncertainties can be compared directly.

The impact of the EIC pseudodata on the experimental uncertainties in the HERAPDF2.0NNLO fits is illustrated in Figs. 3 and 4. Relative uncertainties are shown for the up-valence, down-valence, gluon and total sea-quark densities. The uncertainty bands show the symmetrized total uncertainty, including experimental, model and parametrization contributions,[5] as discussed in detail in [1]. The experimental uncertainties are generally dominant in regions where there is data. A significant reduction in uncertainties for all parton species is observed when adding the EIC pseudodata, as discussed in detail in the following.

---
[4]The hadronic final state invariant mass $W$ is related to the other standard DIS kinematic variables through $W^2 = Q^2(1-x)/x$.

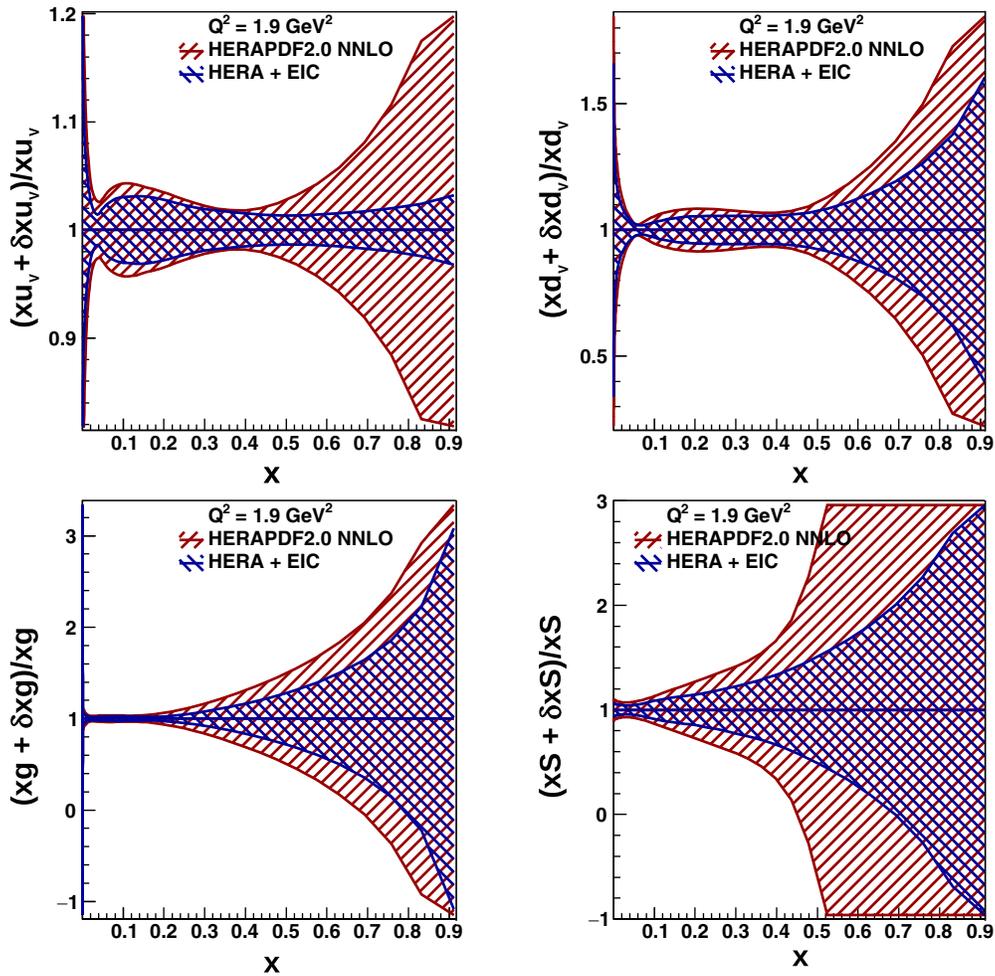

FIG. 3. Impact of simulated EIC data on the NNLO collinear parton distributions of the proton. The bands show relative total uncertainties as a function of $x$ for the up-valence, down-valence, gluon and total sea distributions, for $Q^2 = 1.9$ GeV$^2$. The HERAPDF2.0NNLO total uncertainties (using HERA data alone) are compared with results in which simulated EIC data are also included in the HERAPDF2.0NNLO fitting framework.

Since the high $x$ regime is of particular interest, the potential improvements are most readily visualized on a linear $x$ scale, as shown in Fig. 3 at the starting scale for DGLAP evolution, $Q^2 = 1.9$ GeV$^2$. The EIC impact is most striking in the very large $x$ region $x \gtrsim 0.7$, particularly dramatic for the $u$ valence quark.[6] The improved constraints are traceable to the large integrated luminosity of the EIC pseudodata and the correspondingly improved data precision in the high-$x$ region compared with HERA. The charge-squared weighting of the photon couplings result in stronger sensitivity to the up than to the down quark density. Figure 4 shows the same information as that in Fig. 3, but as a logarithmic function of $x$ and at the electroweak scale, $Q^2 = m_Z^2$ GeV$^2$. At low and intermediate $x$, where our current knowledge of proton collinear structure is dominated by HERA data, the EIC pseudodata continue to have a sizeable impact on the quark densities. While this is in part due to the large EIC luminosity and extended phase-space, the very low $x$ constraints on the valence quarks come primarily from the valence quark number sum rules, and thus derive indirectly from the improved precision at large $x$. The gluon density, which is most strongly constrained by the scaling violations of the neutral current cross sections, is not impacted so strongly, but there is still a notable improvement at intermediate $x$, where the EIC pseudodata improve the precision at low and intermediate $Q^2$.

---

[5]For technical reasons, the number of model and parametrization uncertainties differs between the HERAPDF2.0 baseline and the version with added EIC pseudodata. However, this is expected to cause only comparatively minor differences.

[6]Near to the starting scale for QCD evolution, the PDFs are dominated by the valence quark densities for typically $x \gtrsim 0.2$, with the gluon density becoming dominant at lower $x$ values.

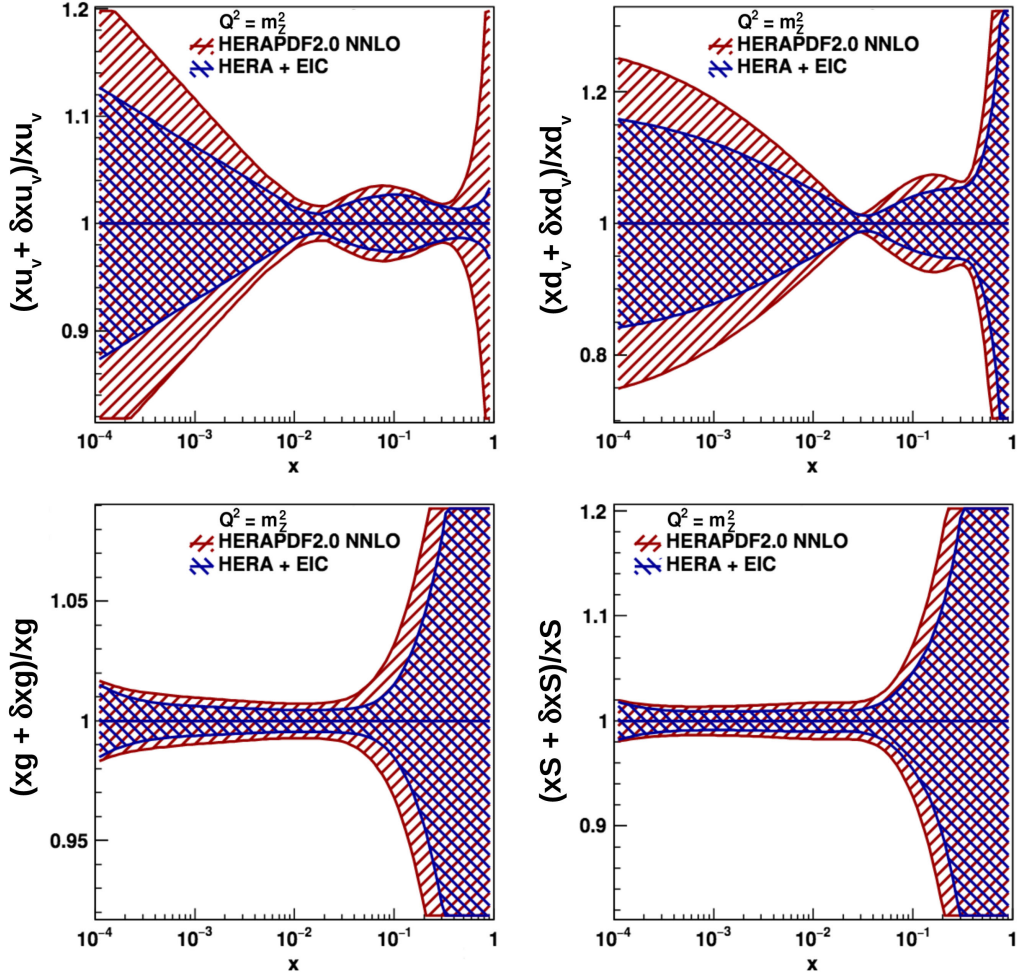

FIG. 4. Impact of simulated EIC data on the NNLO collinear parton distributions of the proton shown on a logarithmic $x$ scale for $Q^2 = m_Z^2$ GeV$^2$. The bands show relative total uncertainties as a function of $x$ for the up-valence, down-valence, gluon and total sea distributions. The HERAPDF2.0NNLO total uncertainties (using HERA data alone) are compared with results in which simulated EIC data are also included in the HERAPDF2.0NNLO fitting framework.

The lowest $x$ and $Q^2$ HERA data show evidence for $\ln(1/x)$ (BFKL) resummation corrections [44]. To investigate whether precise EIC data in the low $x$ region might enhance the overall sensitivity to such effects, modified EIC pseudodata are generated with central values obtained from the NNPDF31sx_nnlo_nllx_as_0118 PDF set [44], which includes the $\ln(1/x)$ resummation effects. Fits at NNLO using the modified pseudodata are compared with and without the addition of next-to-leading $\ln(1/x)$ resummation, implemented using the HELL [45,46] code. The change in the overall $\chi^2$ of the fit between the two cases is negligible, indicating that the EIC data are not sensitive to the BFKL effects. The same conclusion is reached when using a more flexible PDF parametrization [47], designed to give a better description in the small-$x$ region without introducing too many new parameters. While this conclusion could be expected from the smaller access to small $x$ at the EIC compared to HERA, here this conclusion is set on firm grounds for the first time.

### B. Comparison with global PDFs

Several groups worldwide are engaged in "global" fits that constrain the proton PDFs using a range of input data from fixed target DIS experiments, HERA and hadron-hadron colliders, most notably the LHC [48]. Within this global dataset, HERA inclusive DIS data remain a key constraint, in particular at low and intermediate $x$. However, other datasets also play an important role. For example, fixed target DIS data are important in constraining the quarks in the high $x$ region, while LHC measurements are playing an increasingly important role, most notably in the quark flavor decomposition and for the gluon at high $x$ [2,49]. Within such a fit, the impact of any additional data will necessarily be balanced by the pulls of other datasets in the fit. The overall impact is therefore expected to be reduced in comparison to a fit where the EIC pseudodata are added to a more limited, HERA-only, baseline.

The impact of EIC data on the high $x$ PDFs has been studied relative to a recent example global fit, MSHT20 [3,50]. The same cuts, $Q^2 \geq 2$ GeV$^2$ and $W^2 \geq 15$ GeV$^2$ are applied as in [3]. As described in Sec. II the EIC pseudodata are produced using NLO QCD theory and, consistently, with MSHT20NLO PDFs, while the fit is performed at NNLO. This is in order to effectively inject some inconsistency between theory and pseudodata, as one might expect to occur in a real comparison between data and theory. The impact of this procedure on the corresponding PDF uncertainties, relative to the case adopted in Sec. III A where theory and pseudodata agree by construction, is minimal.

As expected, the EIC impact relative to MSHT20 is substantially reduced compared with that relative to HERAPDF2.0. However, there are still significant effects, as illustrated in Fig. 5. Due to the charge-squared coupling of the virtual photon in DIS, up quarks are more strongly impacted than down quarks, such that the biggest impact is on the up-valence distribution, for which a reduction in uncertainties of up to ~50% is observed in the highest $x$ region. The relative impact of data in the low $y$ kinematic region is investigated by comparing the standard selection with results from a dataset restricted to $y > 10^{-2}$. The difference in precision on the up-valence distribution is negligible, which is largely due to the overlapping phase space coverages of the different EIC beam energy configurations, removing the need for difficult measurements at extreme low $y$ values.

The simulated EIC data bring a small, but nonetheless valuable, improvement in the precision on all of the other parton species at all $x$ and $Q^2$ values. The gluon density at the electroweak scale is chosen for illustration here; overall uncertainties are relatively small at such large scales due to the constraints from scaling violations in gluon-initiated DIS.

The results shown here, in the context of the MSHT global PDF fit for the first time, are broadly consistent with previous studies of EIC impact relative to global PDF fits [15–19], though now with pseudodata more representative of expected experimental setups. Some of these previous studies have also noted other smaller impacts in reducing the uncertainties of the strangeness or $d/u$ ratio at high $x$. However there were differences observed between groups and depending on the study [19]. We do not see notable changes in either the strange or $d/u$ ratio uncertainty at high $x$, though our study has only considered proton data and electron beams. The inclusion of deuteron data and/or positron beams would be expected to allow a greater sensitivity to $d$-type quarks and may therefore lead to such improvements.

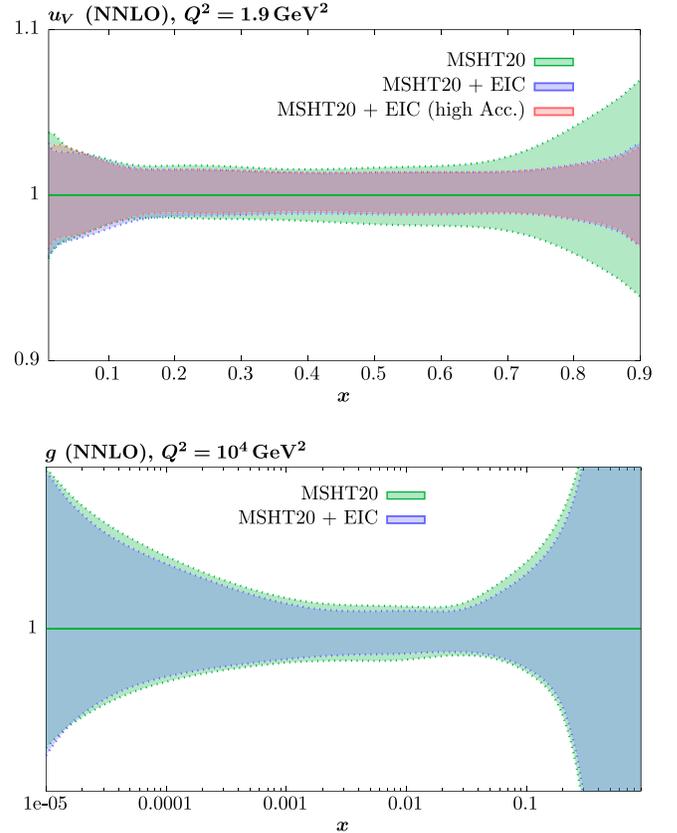

FIG. 5. Impact of simulated EIC data on the collinear proton parton distributions relative to the MSHT20 global fits. The bands show relative uncertainties as a function of $x$, comparing the MSHT20 baseline with results when additionally including EIC data. Top: up valence density at $Q^2 = 1.9$ GeV$^2$, also comparing EIC scenarios with a restriction to $y > 10^{-2}$ (MSHT20 + EIC) with the standard requirement $y > 10^{-3}$ [MSHT20 + EIC (high Acc.)]. Bottom: gluon density at $Q^2 = 10^4$ GeV$^2$.

It is worth noting that the more comprehensive datasets included in global PDF fits bring associated complexities and issues that are not necessarily present in the DIS-only case. As well as the more complex theoretical description of hadron-hadron collisions, and the necessity of including nonperturbative input for example in the nuclear corrections to fixed-target data and hadronization corrections to jet cross sections, there is observed to be a degree of tension between the different datasets that enter the global fit [3,49,51–53]. In the MSHT case, this is accounted for by including an enlarged "tolerance" with respect to textbook statistical expectations when evaluating the corresponding PDF uncertainties [54]. The milder impact of the EIC pseudodata in the MSHT global fit should be considered in this context. In particular, the addition of EIC data may shed light on tensions in other parts of the global fit, giving them value beyond that indicated by the projections shown here.

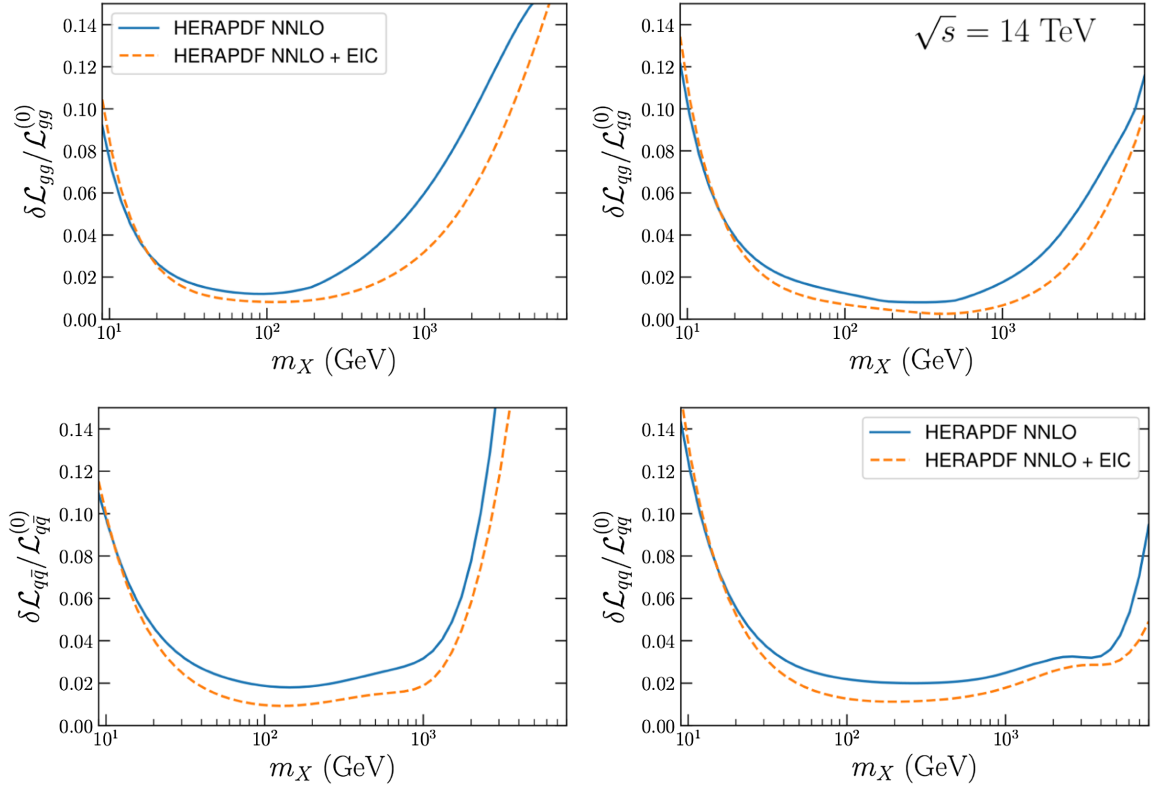

FIG. 6. Impact of simulated EIC data on the LHC collinear PDF luminosities relative to the HERAPDF2.0 PDF fit as a function of the parton-parton invariant mass, $m_X$. Total uncertainties are included. Upper left: gluon-gluon luminosity. Upper right: quark-gluon luminosity. Lower left: quark-antiquark luminosity. Lower right: quark-quark luminosity.

### C. Parton luminosities

Any improvement in PDF uncertainties has consequences for phenomenology at the LHC and this demonstrates the complementarity of the EIC and LHC. The results here have therefore been propagated to investigate the influence on the uncertainties on LHC parton luminosities, which encode the precision with which the initial state conditions are known in $qq$, $gg$, $qg$ and $q\bar{q}$ subprocess collisions as a function of the parton-parton invariant mass, $m_X$. The parton-parton luminosities are calculated as defined in [48].

The impact of the EIC pseudodata on the parton luminosity uncertainties is shown relative to HERAPDF2.0 in Fig. 6. The total uncertainties are shown, incorporating the experimental, model and parametrization components combined in quadrature, as is usual for HERAPDF2.0.[7] All parton luminosities shown have reduced uncertainties following the addition of the EIC pseudodata. The largest impacts are at high $m_X$, reflecting the constraints placed at high $x$ on the PDFs. The gluon-gluon luminosity shows the largest reduction in uncertainty, with effects of up to approximately ∼50% at larger invariant masses. Nonetheless the uncertainty reductions also extend across the invariant mass range down to low masses. This is as expected given the impact of the pseudodata across the whole range of $x$ on the PDFs. The $q\bar{q}$, $qq$, and $qg$ luminosities are similarly reduced in uncertainty down to ∼20 GeV, though again with smaller effects as the invariant mass is reduced.

---

[7]Note however that the model dependence uncertainty eigenvectors are used in pairs here in order to avoid nonsmooth features in the luminosities.

The impact of the EIC pseudodata on the MSHT20 parton luminosity uncertainties (Fig. 7) is significantly milder than that relative to HERAPDF2.0, consistent with the changes seen in the PDF uncertainties. The largest effect is in the gluon-gluon luminosity, which shows a small reduction in uncertainty, approximately constant throughout the invariant mass range, reflecting the reduction in the gluon uncertainties across a range of $x$ values.

A reduction in the gluon-gluon luminosity uncertainty directly affects the precision with which the Higgs production cross-section from gluon-gluon fusion can be predicted [18,19]. The uncertainty in the $gg$ luminosity at $m_H = 125$ GeV reduces from 1.2% to 0.8% with the addition of the EIC pseudodata. The corresponding improvement in precision on the gluon-gluon fusion Higgs production cross section is shown in Fig. 8. The PDF uncertainty, calculated using `ggHiggs` [55–65] with N³LO matrix elements and NNLO PDFs at a central scale of $\mu_f = \mu_r = m_H/2$ and at $\sqrt{s} = 13$ TeV, reduces in proportion to the $gg$ luminosity uncertainty, i.e. also from 1.2% to 0.8%. However, the overall impact on the total uncertainty is much smaller, due to the dominance of the scale uncertainty.

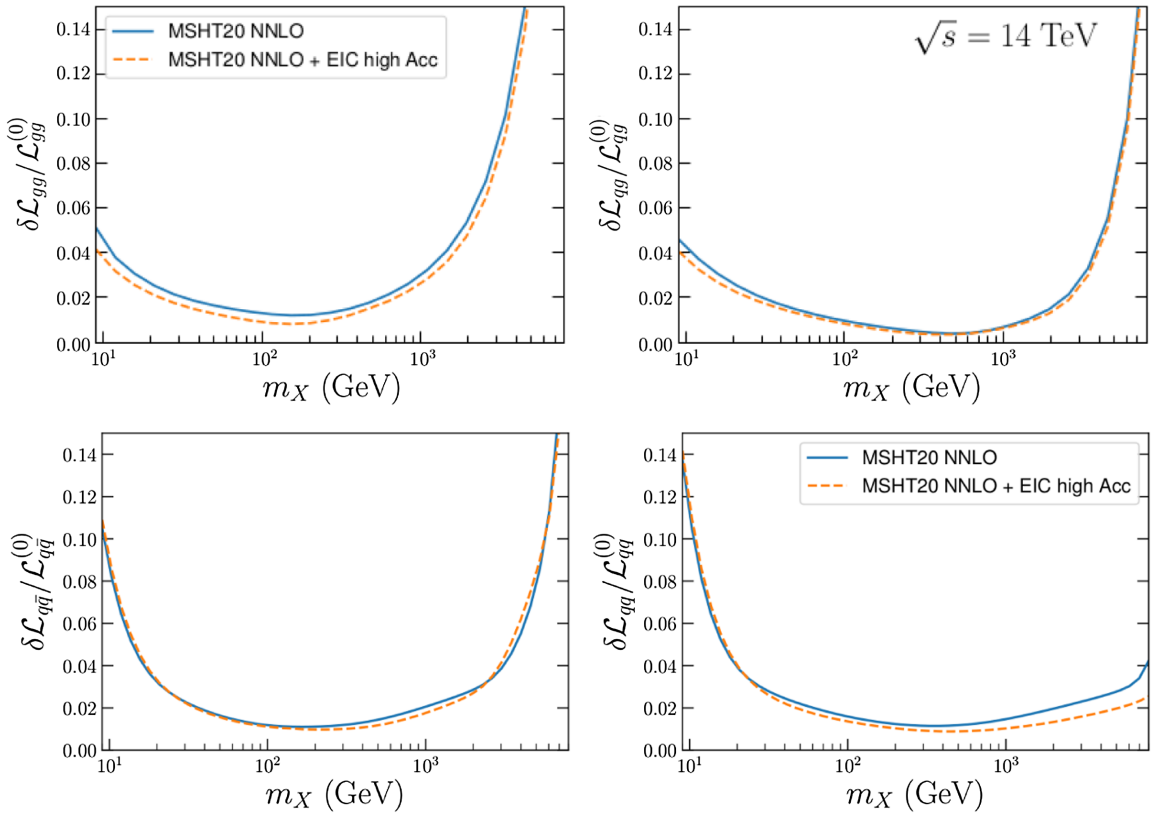

FIG. 7. Impact of simulated EIC data on the LHC collinear PDF luminosities relative to the MSHT20 global fits as a function of the parton-parton invariant mass, $m_X$. Upper left: gluon-gluon luminosity. Upper right: quark-gluon luminosity. Lower left: quark-antiquark luminosity. Lower right: quark-quark luminosity.

There are finite effects for the other parton-parton luminosity cases presented in Fig. 7, with the $qq$ luminosity uncertainty reduced somewhat at larger invariant masses due to the constraints placed by the EIC pseudodata on the large $x$ quark PDFs. The effects on the $qg$ and $q\bar{q}$ luminosities though are small, reflecting the relatively small impact of EIC pseudodata on most of the MSHT20 quark density uncertainties and the integration over rapidities and parton species that must be performed in calculating parton luminosities.

## IV. NUCLEAR PDFs

As the world's first $eA$ collider, the EIC will explore partonic nuclear structure at an unprecedented level of detail. In particular, it opens up a new region at low $x$ that has not been constrained previously in DIS or Drell-Yan data. The partonic structure of nuclei is commonly discussed in terms of nuclear PDFs (nPDFs), or nuclear modification ratios, which encode the deviations of nPDFs from simple scaling of free nucleon PDFs with atomic mass $A$ after appropriately accounting for varying proton-to-neutron ratios using isospin symmetry. The deviations from this scaling with $A$ may be due to binding effects or, at low $x$, to new parton dynamics [66] ('saturation' phenomena) associated with the denser systems of gluons found in heavy nuclei than in nucleons.

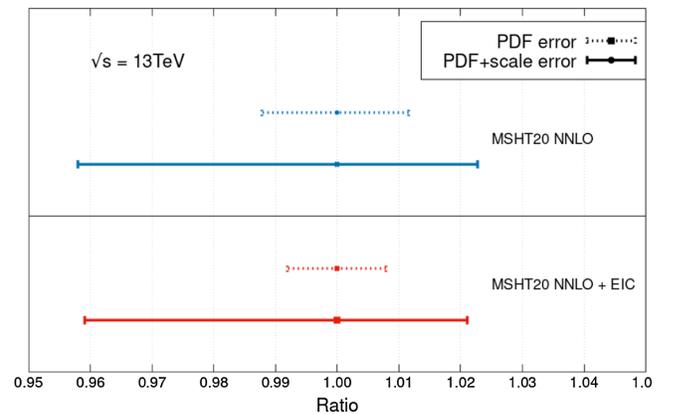

FIG. 8. Impact of simulated EIC data on the Higgs production cross section results via gluon fusion (with $\sqrt{s} = 13$ TeV) at the central scale $\mu = m_H/2$. The cross-section is calculated using N³LO matrix elements but only NNLO PDFs. The dotted lines indicate the PDF only uncertainties, the solid lines are the PDF + scale uncertainties combined in quadrature, with the scale uncertainties determined by varying $\mu$ by a factor of 2 following the 9-point prescription.

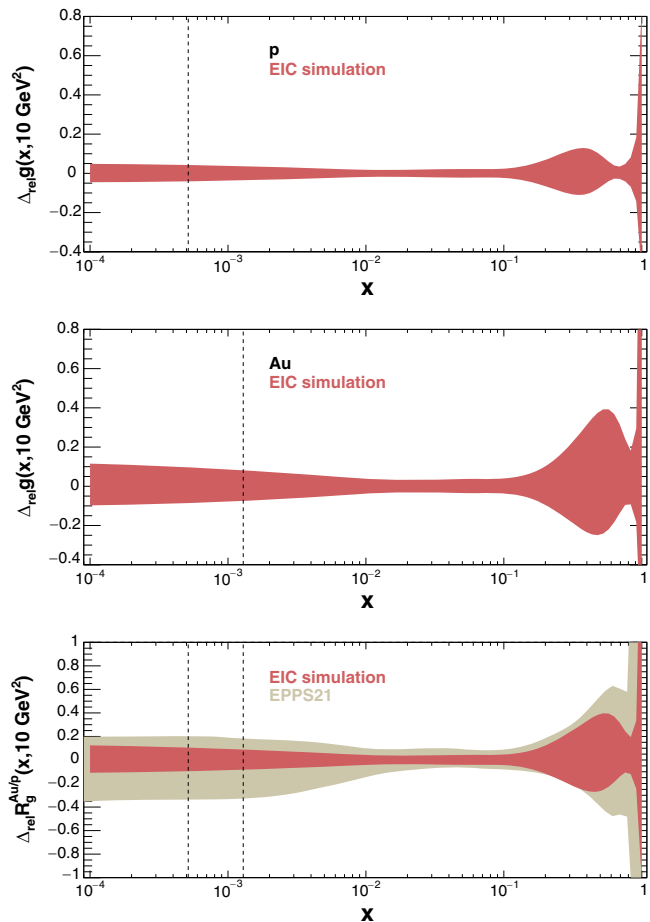

FIG. 9. Impact of EIC data on the understanding of nuclear effects in the collinear gluon distribution, as obtained from DGLAP-based QCD fits. Top: projected relative uncertainty on the gluon density of the proton as a function of $x$ for $Q^2 = 10$ GeV$^2$, using only EIC input data. Middle: projected relative uncertainty on the gluon density of a proton in the gold nucleus as a function of $x$ for $Q^2 = 10$ GeV$^2$, using only EIC input data. Bottom: nuclear modification factors formed from the ratio of projected gluon densities in gold and in the proton. The results obtained using only EIC data are compared with those from a global fit (EPPS21 [10]). Vertical dotted lines indicate the lowest values of $x$ for pseudodata used in the fit, see the text.

Present DIS data feeding into nPDFs are limited to fixed target measurements at large $x$ and relatively low $Q^2$. Data from fixed target and colliding mode hadron-nucleus experiments can be used to extend the sensitivity, but with similar associated theoretical difficulties to those discussed in the proton context in Sec. III. Since the uncertainties in the nuclear modification factors are large in the low $x$ region that will be newly explored in DIS, the EIC is expected to have an impact with relatively modest amounts of $eA$ data.

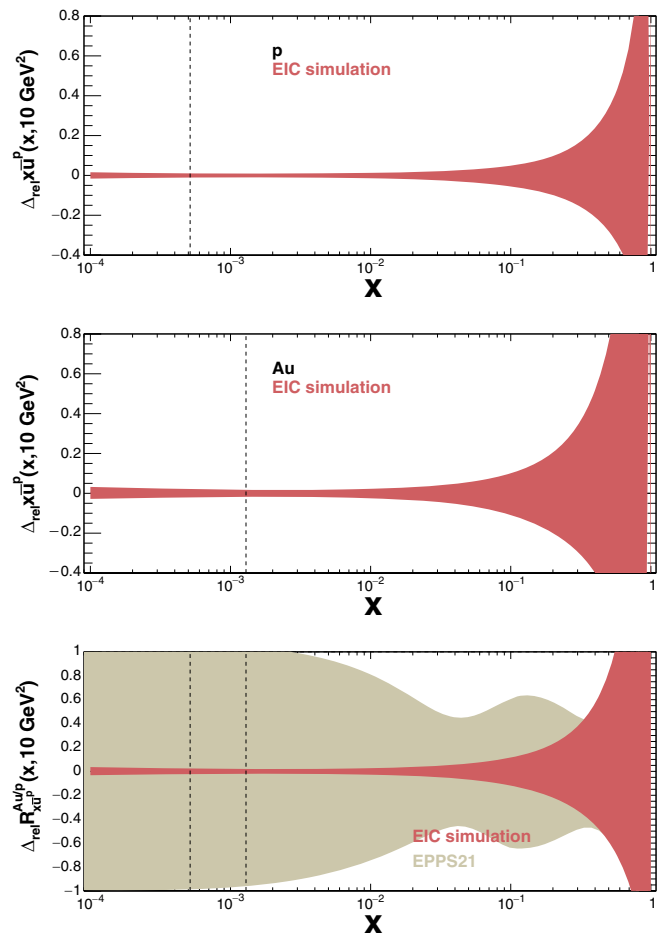

FIG. 10. As for Fig. 9, but for the sea up quark density.

The potential impact on nuclear PDFs of simulated EIC data is studied here in the xFitter framework [43]. Data from EIC only are used as input to fits in which the PDFs evolve according to the next-to-leading order (NLO) DGLAP equations, with a minimum $Q^2$ of 3.5 GeV$^2$ using a parametrization at the starting scale taken from the HERAPDF2.0 studies.[8] Figures 9–11 show the results for the gluon density, the sea up quark density and the up valence quark density, respectively. The relative precision is shown separately for the proton and for gold nuclei, as well as for their ratio, i.e., the nuclear modification factor.[9] The EIC-only projections for the nuclear modification ratios are compared with the precision of a representative current global fit, EPPS21 [10],[10] which includes data from fixed target DIS and Drell-Yan experiments, hard processes in $pA$ collisions at the LHC and $\pi^0$ data from PHENIX.

---

[8]This lower cut in $Q^2$ leads to minimum $x$ values for the pseudodata of 0.0005 and 0.00125 in $ep$ and $eA$, respectively, which are indicated in Figs. 9–11 by vertical dotted lines.

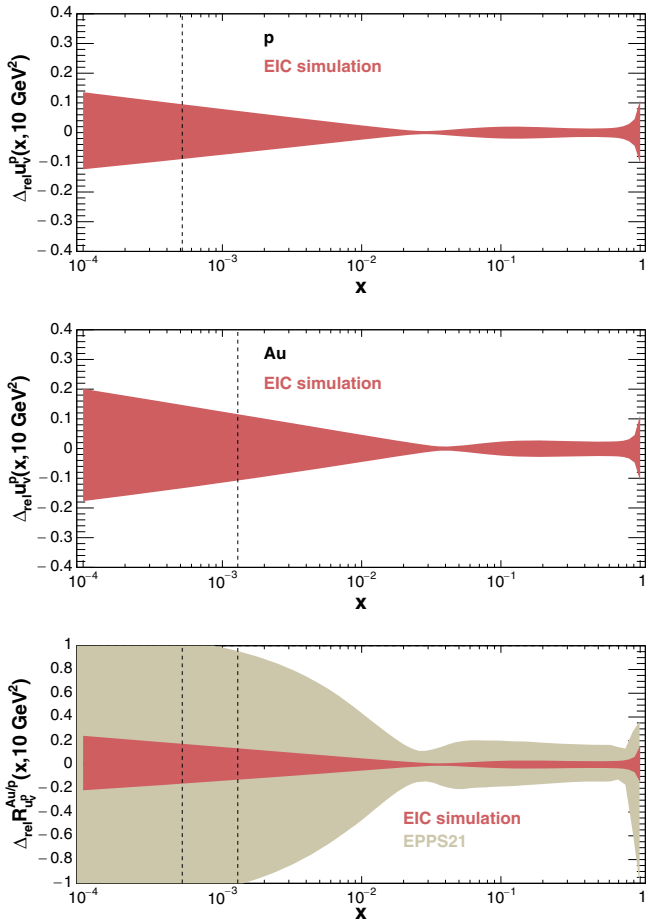

FIG. 11. As for Fig. 9, but for the valence up quark density.

The behavior of gluons at high densities is a cornerstone of the EIC physics program. Since it is most clearly addressed through nuclear dynamics at low $x$, the case of the gluon density (Fig. 9) is pivotal. The precision obtainable using EIC data only is typically at the 5% level for the proton and at the 10% level for gold. Large improvements over the EPPS21 baseline are observed for all $x$ values. The precision in the region $x \sim 0.1$, where there are plentiful input data in the EPPS21 case, is improved by around a factor of 2 when using EIC data. This is partly due to the use of a $\Delta\chi^2 = 1$ condition in defining the uncertainty bands in the EIC case, as would be standard when fitting data from a single experiment, compared with $\Delta\chi^2 = 33$ as applied in the EPPS21 case to account for tensions between the different input datasets.

---

[9]Regions in which the gluon contribution becomes negative, which only appear at low $Q^2$ and small $x$ and are of very little relevance in the EIC kinematics, are removed.

[10]The EPPS21 bands have been computed including both nuclear and proton baseline uncertainties.

The largest improvements in the gluon nuclear modification ratio appear in the previously poorly constrained region below $x \sim 10^{-2}$. Although the EPPS21 fit accesses $x$ values as low as $10^{-5}$, all information below a few times $10^{-3}$ comes from D-meson production in $pPb$ collisions at the LHC, and the uncertainties remain relatively large. The precise and theoretically clean EIC pseudodata extend to approximately $x = 0.001$, opening up an order of magnitude of previously unexplored low $x$ physics in DIS in which novel dynamics may be observed. The gluon nuclear modification ratio is constrained with a precision of approximately 10% in this region. As illustrated in Figs. 10 and 11, even larger, profound, improvements are expected for all other parton species.

In addition to EPPS21, new LHC $pPb$ measurements have also been included in other recent nPDF analyses [9,11–13]. When compared to a previous analysis such as EPPS16 [67], EPPS21 results in a reduction of the uncertainty in the gluon nuclear modification factor by around a factor 2 at small $x$, whereas the reduction is larger in nNNPDF3.0. The variability of the impact of these data between the different analyses suggests that the associated additional theoretical uncertainties and tensions with other datasets have yet to be fully determined. The use of hadron-hadron data in constraining nPDFs also relies on the assumption of collinear factorization [68] for proton-nucleus collisions, which remains to be fully established. In contrast, the EIC will provide a homogeneous and theoretically straightforward dataset from a single experiment and for a single nucleus, with small uncertainties and the possibility of using $\Delta\chi^2 = 1$ in Hessian analyses to quantify the uncertainties.

## V. SUMMARY AND CONCLUSIONS

An analysis studying the impact of simulated EIC inclusive DIS data on the uncertainty of the proton and nuclear PDFs at NNLO and NLO, respectively, is presented.

In the context of a pure-DIS fit for proton PDFs of the HERAPDF2.0 type, the EIC pseudodata which in this study have updated and realistic energy setups, luminosities and detector simulation, offer orders of magnitude larger integrated luminosities than were available at HERA and cover a lower, but still comfortably perturbative, $Q^2$ range at large $x$. They also optimize the kinematic coverage through variations in the beam energies, avoiding the need for measurements at very low $y$. The EIC pseudodata are thus very well suited to improving the precision at large $x$ in a pure DIS fit for the proton PDFs. Their most striking impact occurs for the up-valence quark. Precision PDFs in this large $x$ region are of enormous importance to any hadron-hadron or lepton-hadron experiment with an interest in physics near the $x \to 1$ kinematic limit, notably the energy frontier discovery program at the CERN LHC. These PDF uncertainty reductions at high $x$ propagate into reductions in the parton-parton luminosity uncertainties, with largest effects at large invariant masses.

As expected, smaller improvements in precision are observed when adding EIC pseudodata as an additional component of the global MSHT20 PDF fit. Nonetheless, there are visible improvements for several parton species, the up valence quark once again being the most prominent case. There are also consequent impacts on the parton-parton luminosities, though again notably milder than observed for the HERAPDF2.0 part of the study, with the gluon-gluon luminosity uncertainty reduced relatively uniformly across much of the invariant mass range. The latter then enables a reduction in the PDF uncertainty in gluon-gluon fusion Higgs production, albeit the effect on the combined PDF + scale uncertainty band is smaller. EIC data are also expected to influence fits of this type in a wider manner, enabling a theoretically straightforward approach that may help to unravel tensions between other datasets either due to experimental discrepancies or theoretical complications. In addition, we have studied the sensitivity of EIC inclusive measurements to the presence of large $\ln 1/x$, concluding that the need for resummation of such logarithms, hinted by HERA data, cannot be established with such observables at the EIC.

The potential EIC impact on the nuclear PDFs is investigated in terms of nuclear modification ratios. For the gluon densities large improvements over an EPPS21 baseline are observed for all $x$ values, the largest improvements occurring in the region below $x \sim 10^{-2}$, which has not previously been constrained by DIS data. Even larger improvements are observed for all other parton species.

Altogether, it is clear that the EIC will significantly improve our understanding of the collinear partonic structure of protons and nuclei, in a manner that is achievable using DIS data alone and thus complements the approach of global fits.

## ACKNOWLEDGMENTS

N. A. has received financial support from Xunta de Galicia (Centro singular de investigación de Galicia accreditation 2019-2022, ref. ED421G-2019/05), by European Union ERDF, by the "María de Maeztu" Units of Excellence program MDM2016-0692, and by the Spanish Research State Agency under project No. PID2020–119632GB-I00. This work has been performed in the framework of the European Research Council project No. ERC-2018-ADG-835105 YoctoLHC and the MSCA RISE 823947 "Heavy ion collisions: collectivity and precision in saturation physics" (HIEIC), and has received funding from the European Union's Horizon 2020 research and innovation programme under grant agreement No. 824093. T. C. acknowledges funding from the European Research Council (ERC) under the European Union's Horizon 2020 research and innovation programme (Grant agreement No. 101002090 COLORFREE). This work has been partially funded by the Deutsche Forschungsgemeinschaft (DFG, German Research Foundation)—491245950. L. H.-L. and R. S. T. thank STFC for support via Grant award No. ST/T000856/1.